\documentclass{ws06-procs}

\begin{document}
\title{Statistical analysis of the trigger algorithm for the NEMO project}

\author{Riccio G.$^{1,2}$, D'Angelo G.$^1$, Brescia M.$^{3,2}$}

\address{1 - Department of Physical Sciences, University of Napoli Federico II,
via Cinthia 6, 80126 - Napoli, Italy, E-mail: giuseppe.riccio@na.infn.it\\
2 - INFN - Napoli unit, via Cinthia 6,  80126 - Napoli\\
3 - INAF - Astronomical Observatory of Capodimonte, via Moiariello 16. 80131 - Napoli, Italy}  

\maketitle

\abstracts
{We discuss the performances 
of a trigger implemented for the planned neutrino telescope  NEMO. This trigger
seems capable to discriminate between the signal and the strong background introduced 
by atmospheric muons and by the $\beta$ decay of the $^{40}K$ nuclei present in the water. 
The performances of the trigger, as evaluated on simulated data are analyzed in detail.}

\section{Introduction}

The NEMO (NEutrino Mediterranean Observatory) telescope is designed 
to reveal high and ultra-high energy neutrinos produced by a variety of
astrophysical sources \cite{gr1,gd1}. 
More in particular, the telescope will use a 1 $km^3$ array of PMTs 
to detect the Cherenkov light emitted by  muons produced by a charged 
current interaction of ultra high energy neutrinos. 
The data analysis is made extremely difficult by the presence of a strong
optical background, introduced mainly by atmospheric neutrinos and 
by the $\beta$-decay of the unstable nuclei of
$^{40}$K .
Elsewhere in this proceedings \cite{tr} we have discussed the structure and the 
optimization of the trigger in some detail, here we shall instead focus on 
the evaluation of its performances.

\section{Performances of the trigger algorithm}
As a first step, our analysis required the optimization of the algorithm 
in terms of its three free parameters, namely: sampling time, number 
of datacubes checked, threshold (for further details see \cite{tr}).
In absence of real data, we made extensive use of data simulated with 
the Montecarlo package described in \cite{he} and, more in particular,
we used a set of 1824 simulated events (muon+$^{40}$K) plus 
1200 data containing only the $^{40}$K noise to be used as a 
reference.
(Table \ref{tab:tab1}), 
\begin{table}[h]
\tbl{Values of the parameters used for the optimization of the trigger.
\label{tab:tab1}}
{\begin{tabular}{@{}cccc@{}} \toprule
Sampling Time (ns) & Number of Datacubes checked & Threshold \\
\colrule
5 & 5-10-20-40 & 1 \\
25& 2-3-4-5-6-7-8 & 1 \\
50& 2-3-4 & 1 \\
100 & 1-2 & 1-2\\
300 & 1-2 & 1-2-3\\ \botrule
\end{tabular}}
\end{table}

We then defined two different criteria, the so called Absolute Criterion ($C_{abs}$), 
and Relative Criterion ($C_{rel}$), which provide complementary information. 

\subsection{Absolute Criterion}
In $C_{abs}$, the relative efficiency of the trigger is the sum, 
conveniently normalized, of two terms. The first one is defined as the ratio between 
the number of reconstructed signal PMTs ($N_{PMT}\mu_{ric}$) and the total number of PMTs
($N_{PMT}\mu_{tot}$) which accordingly to the simulation should have
been turned on by the muonic event.
The second term takes instead into account the relative contribution
of PMTs activated by $^{40}$K events ($N_{PMT}K_{ric}$) which are 
erroneously attributed to the
muonic signal:

\begin{equation}
C_{abs} = \frac{100}{2}\!\left[\frac{N_{PMT}\mu_{ric}}{N_{PMT}\mu_{tot}}+\left(1-\frac{N_{PMT}K_{ric}}{N_{PMT}\mu_{ric}+N_{PMT}K_{ric}}\right)\right]
\label{eq:murnf}
\end{equation}

\subsection{Relative Criterion}
$C_{rel}$ generalizes the performances of the algorithm, 
weighing the absolute performances against the ratio between the number of 
PMTs which are recognized as being activated by the muon event and the total
number of PMTs activated in the simulated event($N_{PMT}\mu_{tot}+N_{PMT}K_{tot}$). It needs to be stressed that 
even though this term does not depend on the output of the algorithm, it
estimates the dependence of the performances of the trigger on the specific 
features of the considered event. 
The relative efficiency for $C_{rel}$ is then calculated by adding to $C_{abs}$ 
a term providing the ratio between the number of PMTs activated by muon 
and the totality of PMTs which are turned on by the event. 
This term gives higher weight to the positive feedbacks of the algorithm in the 
case of short tracks (few PMTs are activated):

\begin{eqnarray}{C_{rel}} & = & \frac{100}{2.999}\!\left[ \frac{2}{100} C_{abs}+\left(1-\frac{N_{PMT}\mu_{tot}}{N_{PMT}\mu_{tot}+N_{PMT}K_{tot}}\right)\right]
\end{eqnarray}

\section{Parameters setting}
By applying the analysis criteria just described, we derived the values of 
$C_{abs}$ and $C_{rel}$ for the different configurations listed in Table 1. 
The best combination of parameters is the one that offered the best compromise
among three factors: 
i) the capability to reveal tracks; 
ii) the efficiency in terms of number of PMTs correctly recognized as turned on by events; 
iii) the threshold which needs to be kept as low as possible.  
It turns out that such combination is given by: 
\begin{itemize}
	\item sampling time = 5ns;
	\item number of datacubes checked = 5;
	\item threshold = 1.
\end{itemize}

\section{Results of statistics}

\begin{figure}[ht]
\centerline{\epsfxsize=3in\epsfbox{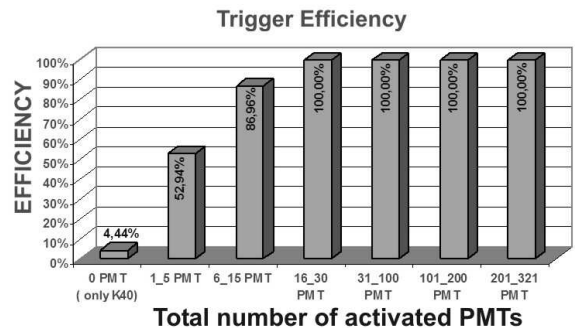}}   
\caption{Efficiency of the trigger for identifying muonic events. \label{fig:efficiency}}
\end{figure}

\begin{figure}[ht]
\centerline{\epsfxsize=3in\epsfbox{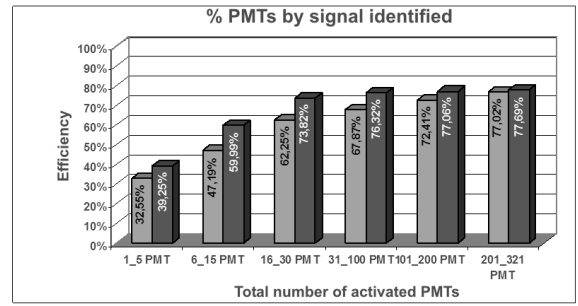}}   
\caption{Efficiency of the trigger for identifying PMTs activated by muonic events. \label{fig:rebuild}}
\end{figure}

Using the above derived triplet of parameters, we estimated the relative 
efficiency of the trigger in terms of its capability of: i) identifying that an event has affected 
a given datastream (Fig. \ref{fig:efficiency}) and, ii) how many PMTs are correctly recognized as turned on by muons
(Fig. \ref{fig:rebuild}).

As it can be seen in Fig. \ref{fig:efficiency} (axis \textbf{y} gives the efficiency of revealing the presence of an event and
the axis \textbf{x} is divided into intervals which indicate the minimum and the maximum number of PMTs activated by a muon in 
different events), the trigger performs well also for short or low energy events. 
In fact in the first interval (0 PMTs, i.e. only $^{40}$K), in the 95.56\% of cases, the trigger flags the absence of muon, 
giving spurious detections only in 4.44\% of cases. 
In the second interval (1-5 PMTs) the trigger flags the presence of muon in the 52.94\% of cases, 
and its efficiency increases with the number of PMTs activated, reaching 100\% above 16 PMTs.

In Fig. \ref{fig:rebuild} we show the "reconstruction" performance of the trigger, i.e. 
how many PMTs activated by a muon correctly identified with respect to the total number 
of PMTs turned on. 
Also in this case the algorithm provides good performances since it 
is capable to reconstruct a minimum of about 33\% PMTs for very low energy 
events up to about 78\% PMTs for high energy ones.

\section{Future development}
In the framework of the NEMO project we foresee the implementation and testing of the trigger
on the real data which will soon be acquired by the prototype tower (4 floors and 16 PMTs) currently 
under deployment and to start its testing on data simulated to match the characteristics of the
first NEMO tower (18 floors and 72 PMTs) which will be 
deployed in a few years.


\begin{thebibliography}{0}

\bibitem{gr1} G. Riccio, 2006, Laurea Thesis University of Napoli Federico II.

\bibitem{gd1} G. d'Angelo, 2006, Laurea Thesis University of Napoli Federico II.
 
\bibitem{tr} G. d'Angelo, G. Riccio \& Brescia M., 2006, {\em Implementation of the trigger algorithm for the NEMO project}, this volume

\bibitem{he} A. Heijboer, Track reconstruction and point source searches with Antares, Universiteit van Amsterdam (2004).

\end{thebibliography}
\end{document}